\documentclass{aa}
\usepackage{txfonts}
\usepackage{graphicx}
\usepackage{float}
\usepackage{natbib}
\usepackage[english]{babel}

\title{Submillimeter number counts at 250~$\mu$m, 350~$\mu$m and 500~$\mu$m in BLAST data}

\author{M. B\'ethermin \and H. Dole \and M. Cousin \and N. Bavouzet}

\institute{Institut d'Astrophysique Spatiale (IAS), Universit\'e Paris-Sud 11 and CNRS (UMR8617), b\^at 121, F-91405 Orsay, France. 
}

\date{Received 18 December 2009 / Accepted 6 March 2010}

\abstract{The instrument BLAST (Balloon-borne Large-Aperture Submillimeter
  Telescope) performed the first deep and wide extragalactic
  survey at 250, 350 and 500 $\mu$m. The extragalactic number counts
  at these wavelengths are important constraints for modeling the evolution of infrared galaxies.}
{We estimate the extragalactic number counts in the
  BLAST data, which allow a comparison with the results of the P(D)
  analysis of Patanchon et al. (2009). }
{ We use three methods to identify the submillimeter sources. 1) Blind
  extraction using an algorithm when the observed field is confusion-limited and another one when the observed field is instrumental-noise-limited. The photometry is computed with a new
  simple and quick point spread function (PSF) fitting routine (FASTPHOT). We use Monte-Carlo
  simulations (addition of artificial sources) to characterize the
  efficiency of this extraction, and correct the flux boosting and the
  Eddington bias. 2) Extraction using a prior. We use the
  \textit{Spitzer} 24 $\mu$m galaxies as a prior to probe slightly
  fainter submillimeter flux densities. 3) A stacking
  analysis of the \textit{Spitzer} 24 $\mu$m galaxies in the BLAST
  data to probe the peak of the differential submillimeter counts.  }
{ With the blind extraction, we reach 97~mJy, 83~mJy and 76~mJy at 250~$\mu$m, 350~$\mu$m and 500~$\mu$m respectively with a 95\% completeness. With the prior extraction, we reach 76~mJy, 63~mJy, 49~mJy at 250~$\mu$m, 350~$\mu$m and 500~$\mu$m respectively. With the stacking analysis, we reach 6.2~mJy, 5.2~mJy and 3.5~mJy at 250 $\mu$m, 350~$\mu$m and 500~$\mu$m respectively. The differential submillimeter number counts are derived, and start showing a turnover at flux densities decreasing with increasing wavelength.}
{ There is a very good agreement with the P(D) analysis of Patanchon
  et al. (2009). At bright fluxes ($>$100~mJy), the Lagache et
  al. (2004) and Le Borgne et al. (2009) models slightly overestimate
  the observed counts, but the data agree very well near the
  peak of the differential number counts. Models predict that the galaxy
  populations probed at the peak are likely $z\sim 1.8$ ultra-luminous
  infrared galaxies.}

\titlerunning{BLAST extragalactic number counts}
\authorrunning{B\'ethermin et al.}

\begin{document}

\maketitle

\section{Introduction}

Galaxy number counts, a measurement of the source surface density as a
function of flux density, are used to evaluate the global evolutionary
photometric properties of a population observed at a given
wavelength. These photometric properties mainly depend on the source
redshift distribution, spectral energy distribution (SED), and
luminosity distribution in a degenerate way for a given
wavelength. Even though this is a rather simple tool, measurements of number
counts at different observed wavelengths greatly help in constraining
those degeneracies. Backward evolution models, among these \citet{Chary2001,Lagache2004,Gruppioni2005,Franceschini2009,Le_Borgne2009,Pearson2009,Rowan2009,Valiante2009}
are able to broadly reproduce (with different degrees of accuracy) the
observed number counts from the near-infrared to the millimeter
spectral ranges, in addition to other current constraints, like such as
measured luminosity functions and the spectral energy distribution of
the Cosmic Infrared Background (CIB)
\citep{Puget1996,Fixsen1998,Hauser1998,Lagache1999,Gispert2000,Hauser2001,Kashlinsky2005,Lagache2005,Dole2006}.
In the details, however, the models disagree in some aspects like the
relative evolution of luminous and ultra-luminous infrared galaxies
(LIRG and ULIRG) and their redshift distributions, or the mean
temperature or colors of galaxies, as is shown for instance in
\citet{Le_Floch2009} from Spitzer 24~$\mu$m deep observations.

One key spectral range lacks valuable data to get accurate
constraints as yet: the sub-millimeter range, between 160~$\mu$m and 850~$\mu$m,
where some surveys were conducted on small areas. Fortunately this
spectral domain is intensively studied with the BLAST
balloon experiment \citep{Devlin2009} and the \textit{Herschel} and \textit{Planck}
space telescopes. This range, although it is beyond the maximum of the
CIB's SED in wavelength, allows us to constrain the poorly-known cold
component of galaxy SED at a redshift greater than a few
tenths. Pioneering works have measured the local luminosity function
\citep{Dunne2000} and shown that most milli-Jansky sources lie at
redshifts $z>2$
\citep{Ivison2002,Chapman2003a,Chapman2005,Ivison2005,Pope2005,Pope2006}. Other
works showed that the galaxies SED selected in the submillimeter range
\citep{Benford1999,Chapman2003b,Sajina2003,Lewis2005,Beelen2006,Kovacs2006,Sajina2006,Michalowski2009}
can have typically warmer temperatures and higher luminosities than
galaxies selected at other infrared wavelengths.

Data in the submillimeter wavelength with increased sensitivity are thus needed
to match the depth of infrared surveys, conducted by Spitzer in the
mid- and far-infrared with the MIPS instrument \citep{Rieke2004} at
24~$\mu$m, 70~$\mu$m and 160~$\mu$m
\citep{Chary2004,Marleau2004,Papovich2004,Dole2004,Frayer2006a,Frayer2006b,Rodighiero2006,Shupe2008,Frayer2009,Le_Floch2009,Bethermin2010a} as well as the near-infrared range with the IRAC instrument
\citep{Fazio2004a} between 3.6~$\mu$m and 8.0~$\mu$m
\citep{Fazio2004b,Franceschini2006,Sullivan2007,Barmby2008,Magdis2008,Ashby2009}.
Infrared surveys have allowed the resolution of the CIB by identifying
the contributing sources -- directly at 24~$\mu$m and 70~$\mu$m, or
indirectly trough stacking at 160~$\mu$m \citep{Dole2006,Bethermin2010a}.

Although large surveys cannot solve by themselves all the unknowns
about the submillimeter SED of galaxies, the constraints given by the
number counts can greatly help in unveiling the statistical SED shape
of submillimeter galaxies as well as the origin of the submillimeter
background.

The instrument BLAST (Balloon-borne Large-Aperture Submillimeter Telescope,
\citet{Pascale2008}) performed the first wide and deep survey in the
250-500~$\mu$m range \citep{Devlin2009} before the forthcoming
\textit{Herschel} results. \citet{Marsden2009} show that sources detected by
\textit{Spitzer} at 24 $\mu$m emit the main part of the submillimeter
background. \citet{Khan2009} claimed that only 20\% of the CIB is resolved by the sources brighter than 17~mJy at 350~$\mu$m. \citet{Patanchon2009} has performed a $P(D)$ fluctuation
analysis to determine the counts at BLAST wavelength (250~$\mu$m, 350~$\mu$m and 500~$\mu$m). In this paper we propose another method to estimate the
number counts at these wavelengths and compare the results with those of \citet{Patanchon2009}.

\begin{figure}
\centering
\includegraphics[width=8cm]{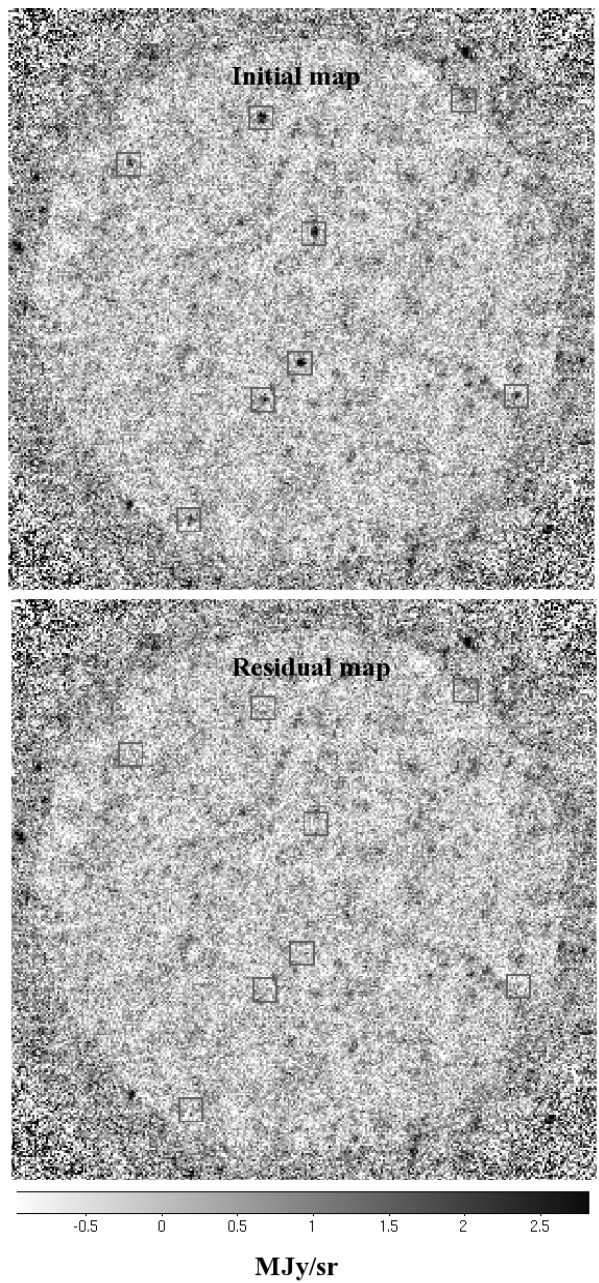}
\caption{\label{fig:figext} Position of sources brighter than the 95\% completeness flux at 250~$\mu$m in deep zone. \textit{Top}: initial map. \textit{Bottom}: residual map. The area out of the mask are represented darker. These 1$^\circ \times $1$^\circ$ map are centered on the coordinates (RA,Dec) = (3h32min30s,-27$^\circ$50'). The horizontal axis is aligned with the right ascension.}
\end{figure}

\section{Data}

\subsection{BLAST sub-millimeter public data in the Chandra Deep Field South (CDFS)}

The BLAST holds a bolometer array, which is the precursor of the spectral and photometric imaging
receiver (SPIRE) instrument on \textit{Herschel}, at the focus of a
1.8~m diameter telescope. It observes at 250~$\mu$m, 350~$\mu$m and 500~$\mu$m, with a 36'', 42'' and 60'' beam, respectively \citep{Truch2009}. 

An observation of the Chandra Deep Field South (CDFS) was performed
during a long duration flight in Antarctica in 2006, and the data of
the two surveys are now public: a 8.7~deg$^2$ shallow field and a 0.7~deg$^2$ confusion-limited \citep{Dole2004} field in the center part of the first one. We use the non-beam-smoothed maps and associated
point spread function (PSF) distributed on the BLAST
website\footnote{http://www.blastexperiment.info}. The signal and noise maps were generated by the SANEPIC algorithm \citep{Patanchon2008}.

\subsection{\textit{Spitzer}  24~$\mu$m data in the CDFS}

Several infrared observations were performed in the CDFS. The
\textit{Spitzer} Wide-Field InfraRed Extragalactic (SWIRE) survey
overlaps the CDFS BLAST field at wavelengths between 3.6~$\mu$m and
160~$\mu$m. We used only the 24~$\mu$m band, which is 80\%
complete at 250$\mu$Jy. The completeness is defined as the probability to find a source of a given flux in a catalog. The Far-Infrared Deep Extragalactic Legacy
(FIDEL) survey is deeper but narrower (about 0.25~deg$^2$) than
SWIRE and 80\% complete at 57~$\mu$Jy at 24 $\mu$m. We used the
\citet{Bethermin2010a} catalogs constructed from these two
surveys. These catalogs were extracted with SExtractor
\citep{Bertin1996} and the photometry was performed with the allstar
routine of the DAOPHOT package \citep{Stetson1987}. The completeness of this catalog was characterized with Monte-Carlo
simulations (artificial sources added on the initial map and
extracted).

\begin{figure}
\centering
\includegraphics{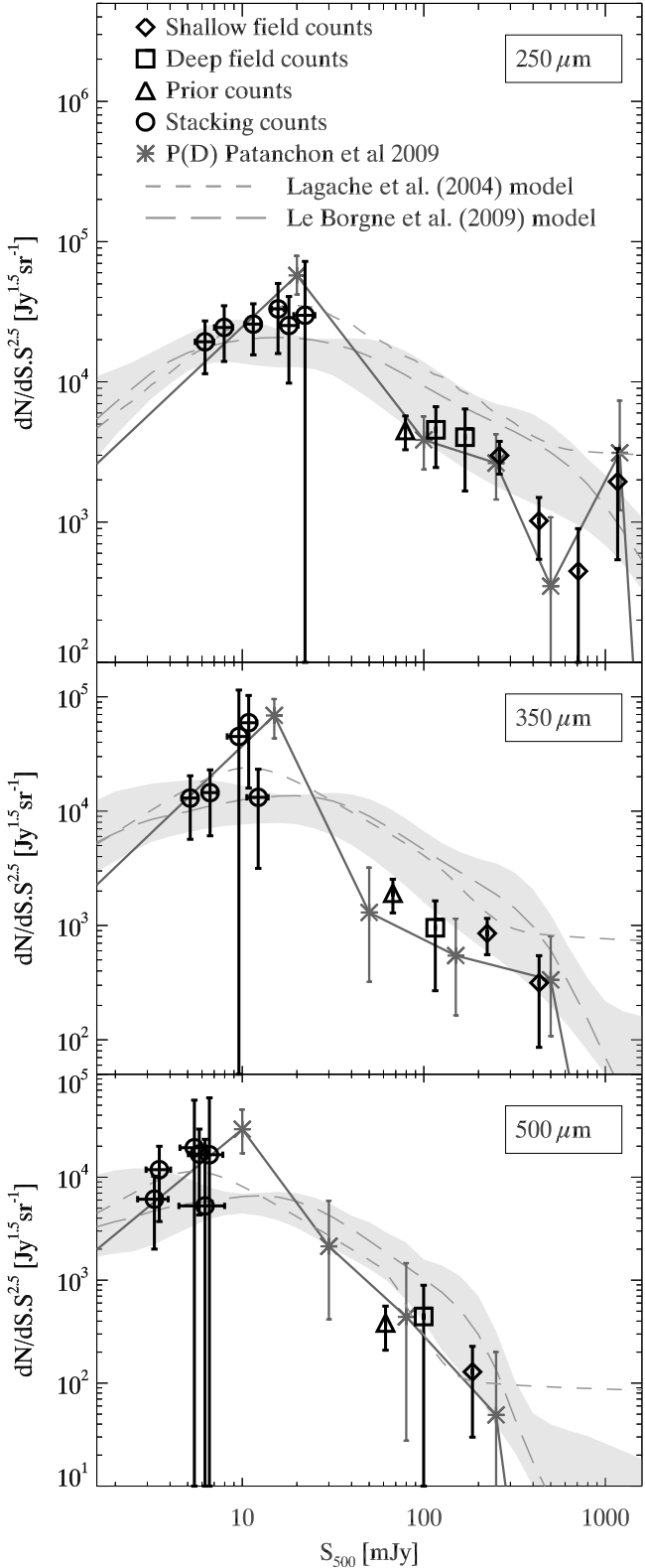}
\caption{
  \label{fig:nc} Extragalactic number counts at 250 $\mu$m in the BLAST
  data. \textit{diamond}: counts deduced from the source catalog on
  the whole shallow field; \textit{square}: counts deduced from the
  catalog of the deep part of the field; \textit{triangle}: counts
  deduced from catalog of the deep part of the field with a 24
  $\mu$m prior (this measurement gives only a lower limits to the
  counts); \textit{cross}: counts computed with a stacking analysis;
  \textit{grey asterisk}: counts computed with a P(D) analysis by
  \citet{Patanchon2009}; \textit{grey short dashed line}: \citet{Lagache2004}
  model prediction; \textit{grey long dashed line and grey area}:
  \citet{Le_Borgne2009} model prediction and 1-$\sigma$ confidence area.}
\end{figure}

\section{Blind source extraction and number counts}
\label{section:blind}

We started with a blind source extraction in the BLAST bands. Each
wavelength was treated separately. For each wavelength we defined two masks: a shallow zone (about 8.2 deg$^2$) covering the whole field except the noisier edge; and a deep zone (about 0.45 deg$^2$) in the
center of the confusion-limited area. We used different extraction
methods in the shallow zone and the deep one, but the photometry and
the corrections of the extraction bias were the same.

\subsection{Detector noise-limited extraction (shallow zone)}

\label{section:noise_ext}

In the shallow zone we used the non-smoothed map and the corresponding
map of the standard deviation of the noise. The map was then cross-correlated
by the PSF. The result of this cross-correlation is
\begin{equation}
m_{conv}(i_{0},j_{0}) = \sum_{i=-N}^{+N} \sum_{j=-N}^{+N} m(i_0+i,j_0+j) \times PSF(i,j),
\end{equation}
where $m_{conv}(i_{0},j_{0})$ is the flux density in the pixel
$(i_0,j_0)$ of the cross-correlated map, $m(i,j)$ the flux density in the
pixel $(i,j)$ of the map, and $PSF(i,j)$ the value of the normalized PSF in the pixel $(i,j)$ (the center of the PSF is in the
center of the pixel $(0,0)$). The PSF size is (2N+1)$\times$(2N+1)
pixels. The standard deviation of the noise in the cross-correlated map is
thus
\begin{equation}
n_{conv}(i_{0},j_{0}) = \sqrt{\sum_{i=-N}^{+N} \sum_{j=-N}^{+N} n^2(i_0+i,j_0+j) \times PSF^2(i,j)},
\end{equation}
where $n$ ($n_{conv}$) is the initial (cross-correlated) map of
the standard deviation of the noise.

We found the pixels where $m_{conv}/n_{conv}>3$ and kept the local
maxima. The precise center of the detected sources was computed by a
centroid algorithm. This low threshold caused lots of spurious detections, but helped to deblend the fluxes of 3 to 4-sigma sources and avoided to overestimate their fluxes. We could thus limit the flux boosting effect. A final cut in flux after the PSF fitting photometry eliminated the main part of these sources. We performed the extraction algorithm on the flipped map (initial map multiplied by a factor of -1) to check it. We found few spurious sources brighter than the final cut in flux determined in the Sect. \ref{section:completeness}. We found a spurious rate of 12\%, 11\% and 25\% at 250~$\mu$m, 350~$\mu$m and 500~$\mu$m, respectively.

\subsection{Confusion-limited extraction (deep zone)}

\label{section:conf_ext}

In the confusion-limited zone we also used a non-smoothed map.  In
this region the noise is dominated by the confusion and not by the
instrumental noise. Consequently, the method based on instrumental
noise presented in the Sect. \ref{section:noise_ext} is not
relevant. We used an atrou wavelet filtering \citep{Starck1999,
  Dole2001} to remove fluctuations at scales larger than 150''. Then we divided the resulting map by $\sigma_{filtered~map}$, which is the
standard deviation of the pixel values on the filtered map in the working area. We
finally kept local maxima with a signal greater than 3. The center of
the sources was also determined by a centroid algorithm. The initial
map and the cleaned map are shown in Fig. \ref{fig:figext}. When we flip the map, we find no spurious source brighter than the final cut in flux determined in Sect. \ref{section:completeness}.

\subsection{A simple and quick PSF fitting routine: FASTPHOT}

\label{section:psffit}

For both noise- and confusion-limited extraction, we apply the
same quick and simple PSF fitting routine on the non-beam-smoothed map. This
routine fits all the detected sources at the same time and is
consequently efficient for deblending (although no source was
  blended in this case; but source-blending will be an issue for an extraction using a prior, detailed in
  Sect. \ref{section:prior}. We suppose that the noise is Gaussian
and the position of sources is known. We then maximize the likelihood
\begin{equation}
L(m|S) = \prod_{pixels} C(n) \times exp\Bigl[ -\frac{\bigl ( m-\sum_{i=1}^{Nsources} PSF_{x_i, y_i} \times S_i \bigl )^2}{2 n^2} \Bigl],
\end{equation}
where $m$ and $n$ are the map and the noise map. $PSF_{x_i, y_i}$ is a unit-flux
PSF centered at the position $(x_i, y_i)$, which are the coordinates
of the i-th source. These coordinates are not necessarily
integers. $C(n)$ is a normalization constant and depends only of the
value of the noise map. S is a vector containing the flux of the
sources.

The value of S, which maximizes the likelihood, satisfies the
  following linear equation stating that the derivative of the
  likelihood logarithm equals zero
\begin{equation}
	\forall i, ~ 0 = \frac{\partial{log\bigl(L(m|S)\bigl)}}{\partial S_i} = A.S + B,
\end{equation}
where A is a matrix and B a vector defined by
\begin{equation}
A = (a_{ij}) = - \sum_{pixels} \frac{PSF_{x_i,y_i} \times PSF_{x_j,y_j}}{n^2}
\end{equation}
\begin{equation}
B = (b_i) = \sum_{pixels} \frac{PSF_{x_i,y_i} \times map}{n^2}.
\end{equation}
To perform this operation fast, we used a 70''$\times$70''
(respectively 90''$\times$90'' and 110''$\times$110'') PSF at 250 $\mu$m
(respectively 350~$\mu$m and 500~$\mu$m). This PSF, provided by the BLAST
  team, is the response for a unit-flux source and takes into account
  all the filtering effects. We used the conjugate gradient method to
solve this equation quickly.

This routine was tested with 200$\times$200 pixels
  simulated maps containing 400 sources at a known positions with a
  beam of 10 pixels FWHM. The flux of all sources was perfectly
  recovered in the case where no noise was added. This routine
(FASTPHOT) performs simultaneous PSF fitting photometry of 1000 sources
in less than 1 second. It is publicly available\footnote{on the IAS
  website http://www.ias.u-psud.fr/irgalaxies/}.

\subsection{Completeness and photometric accuracy}

\label{section:completeness}

The completeness is the probability to detect a source of a given flux
density. We measured it with a Monte-Carlo simulation. We added
artificial point sources (based on PSF) on the initial map at random
positions and performed the same source extraction and photometry
algorithm as for the real data. A source was considered to be detected
if there was a detection in a 20'' radius around the center of the
source. Table \ref{tab:comp} gives the 95\% completeness flux
density (for which 95\% of sources at this flux are detected)
for different wavelengths and depths.

The photometric noise was estimated with the scatter of the recovered
fluxes of artificial sources. We computed the standard deviation of the
difference between input and output flux. This measurement includes
instrumental and confusion noise ($\sigma_{tot} =
\sqrt{\sigma_{instr}^2+\sigma_{conf}^2}$). The results are given in Table
\ref{tab:comp}. In the deep area, the photometric uncertainties are
thus dominated by the confusion noise. The estimations of the
confusion noise between the deep and shallow areas are consistent. It
shows the accuracy and the consistency of our method. 

Note that the uncertainties on flux densities in the \citet{Dye2009}
catalog (based only on instrumental noise) are consequently largely
underestimated in the confusion-limited area. Indeed, their 5
$\sigma$ detection threshold (based only on instrumental noise) at 500
$\mu$m in the deep zone corresponds to 1.76 $\sigma$ if we also include the confusion noise.

The faint flux densities are overestimated due to the classical flux boosting effect, . This bias was measured for all bands for 60 flux densities between 10 mJy and 3 Jy with the results of the Monte-Carlo
simulations. The measured fluxes were deboosted with this relation. We
cut the catalogs at the 95\% completeness flux, where the boosting
factor is at the order of 10\%. Below this cut, the boosting effect increases too quickly to be safely corrected. We also observed a little underestimation at high flux of 1\%, 0.5\% and 0.5\% at 250~$\mu$m, 350~$\mu$m and 500~$\mu$m. It is due to FASTPHOT, which assumes that the position is perfectly known, which is not true, especially for a blind extraction.

\begin{table*}
\centering
\begin{tabular}{|l|rr|rr|rr|rr|}
\cline{2-9}
\multicolumn{1}{c|}{} & \multicolumn{2}{|c|}{95\% completeness} & \multicolumn{2}{|c|}{instrumental noise} & \multicolumn{2}{|c|}{total photometric noise} & \multicolumn{2}{|c|}{deduced confusion noise} \\
\cline{2-9}
\multicolumn{1}{c|}{} & \multicolumn{2}{|c|}{mJy} &  \multicolumn{2}{|c|}{mJy} &  \multicolumn{2}{|c|}{mJy} &  \multicolumn{2}{|c|}{mJy}\\
\cline{2-9}
\multicolumn{1}{c|}{} & shallow &deep & shallow & deep & shallow & deep & shallow & deep\\
\hline
250 $\mu$m & 203 & 97 &  37.7 & 11.1 & 47.3 & 24.9 & 28.6 & 22.3\\
350 $\mu$m & 161 & 83 & 31.6 & 9.3 & 35.8 & 20.3 & 16.8 & 18.0\\
500 $\mu$m & 131 & 76 &  20.4 & 6.0 & 26.4 & 17.6 & 16.7 & 16.5\\
\hline
\end{tabular}
\caption{\label{tab:comp} 95\% completeness flux density and photometric noise for different depths at different wavelengths. The instrumental noise is given by the noise map. The total photometric noise includes the instrumental and confusion noise and is determined by Monte-Carlo simulations. The confusion noise is computed with the formula $\sigma_{conf} = \sqrt{\sigma_{tot}^2-\sigma_{instr}^2}$.}
\end{table*}

\subsection{Number counts}

We computed number counts with catalogs corrected for boosting. For
each flux density bin we subtracted the number of spurious detections estimated in the Sects. \ref{section:noise_ext} and \ref{section:conf_ext} to the number of detected sources and divided the number of sources by the size of the bin, the size of the field and the completeness.

We also applied a corrective factor for the Eddington bias. We assumed a
distribution of flux densities in $dN/dS \propto S^{-r}$ with
$r=3\pm0.5$. This range of possible values for r was estimated
considering the \citet{Patanchon2009} counts and the \citet{Lagache2004} and
\citet{Le_Borgne2009} model predictions. We then randomly kept sources
with a probability given by the completeness and added a random
Gaussian noise to simulate photometric noise. Finally we computed the
ratio between the input and output number of sources in each bin. We
applied a correction computed for $r = 3$ to each point. We estimated
the uncertainty on this correction with the difference between
corrections computed for $r = 2.5$ and $r = 3.5$. This uncertainty was
quadratically combined with a Poissonian uncertainty (clustering
effects are negligible due to the little number of sources in the
map, see appendix \ref{section:cluscounts}).

The calibration uncertainty of BLAST is 10\%, 12\% and 13\% at 250~$\mu$m, 350~$\mu$m and 500~$\mu$m respectively \citep{Truch2009}. This uncertainty is combined with other uncertainties on the counts. The results are plotted in Fig. \ref{fig:nc} and given in Table \ref{tab:nc} and interpreted in
Sect. \ref{section:counts_int}.

\subsection{Validation}

We used simulations to validate our method. We generated 50 mock
catalogs based on the \citet{Patanchon2009} counts, and which covered 1 deg$^2$ each. These sources are spatially homogeneously distributed. We then generated the associated maps
at 250 $\mu$m. We used the instrumental PSF, and added a
gaussian noise with the same standard deviation as in the deepest part
of real map.

\begin{figure}[hb]
\centering
\includegraphics{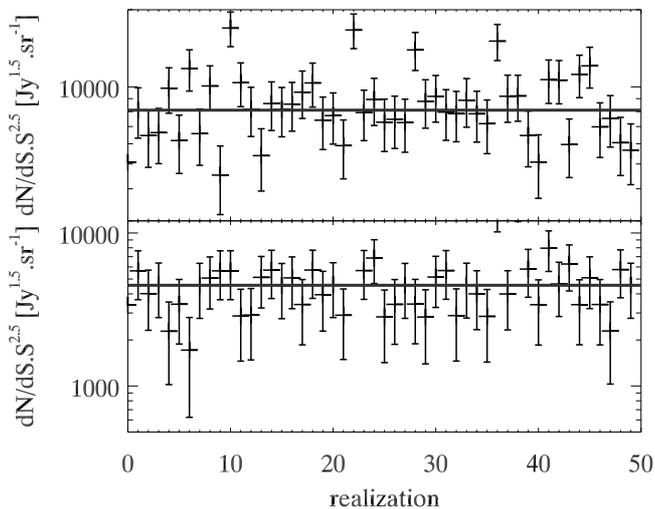}
\caption{\label{fig:valid_cla} Number counts at 250 $\mu$m deduced
  from a blind extraction for 50 realizations of a simulation based on
  the \citet{Patanchon2009} counts. The horizontal \textit{solid line} represents the
  input count value. The lower panel is the result of the 100-141~mJy
  bin, the upper panel is the 141-200~mJy bin.}
\end{figure}

We performed an extraction of sources and computed the number counts with
the method used in the confusion limited part of the field (Sect.
\ref{section:conf_ext}). We then compared the output counts with the
initial counts (Fig. \ref{fig:valid_cla}). We used two flux density
bins: 100-141~mJy and 141-200~mJy. We found no significant bias. The
correlation between the two bins is 0.46. The neighbor points
  are thus not anti-correlated as in the \citet{Patanchon2009} P(D)
  analysis.

The same verification was done on 20 \citet{Fernandez-Conde2008}
simulations (based on the \citet{Lagache2004} model). These
simulations include clustering. This model overestimates the number of
the bright sources, and the confusion noise is thus stronger. The 95\% completeness is then reach at 200~mJy. But there is also a very good agreement between input and output counts in bins brighter
than 200~mJy. We found a correlation between two first bins of 0.27.

\section{Source extraction using \textit{Spitzer} 24 $\mu$m catalog as a prior}

\label{section:prior}

In addition to blind source extraction in the BLAST data (Sect.~\ref{section:blind}) we also performed a source extraction using a prior.

\subsection{PSF fitting photometry at the position of the \textit{Spitzer} 24~$\mu$m}

\label{section:prior_cat}

  The catalogs of infrared galaxies detected by \textit{Spitzer} contain
more sources than the BLAST catalog. The 24~$\mu$m Spitzer PSF has a
Full Width at Half Maximum (FWHM) of 6.6''. It is smaller than the BLAST PSF (36'' at
  250~$\mu$m). Consequently, the position of the Spitzer sources is
known with sufficient accuracy when correlating with the BLAST data.

We applied the FASTPHOT routine (Sect. \ref{section:psffit}) at the positions of 24~$\mu$m sources. We used the \citet{Bethermin2010a} SWIRE catalog cut at $S_{24}=250~\mu$Jy (80\% completeness). In order to avoid software instabilities, we kept in our analysis only the brightest \textit{Spitzer} source in a 20'' radius area (corresponding to 2 BLAST pixels). The corresponding surface density is 0.38, 0.49 and 0.89 Spitzer source per beam\footnote{the beam solid angles are taken as 0.39~arcmin$^2$, 0.50~arcmin$^2$ and 0.92~arcmin$^2$ at 250~$\mu$m, 350~$\mu$m and 500~$\mu$m respectively.} at 250~$\mu$m, 350~$\mu$m and 500~$\mu$m, respectively.
  
This method works only if there is no astrometrical offset between the input 24~$\mu$m catalog and the BLAST map. We stacked the BLAST sub-map centered on the brightest sources of the SWIRE catalog and measured the centroid of the resulting artificial source. We found an offset of less than 1''. It is negligible compared to the PSF FWHM (36'' at 250~$\mu$m).

  We worked only in the central region of the deep confusion-limited
  field (same mask as for blind extraction), where the
  photometric noise is low.

\subsection{Relevance of using \textit{Spitzer} 24~$\mu$m catalog as a prior}

\begin{figure}
\centering
\includegraphics{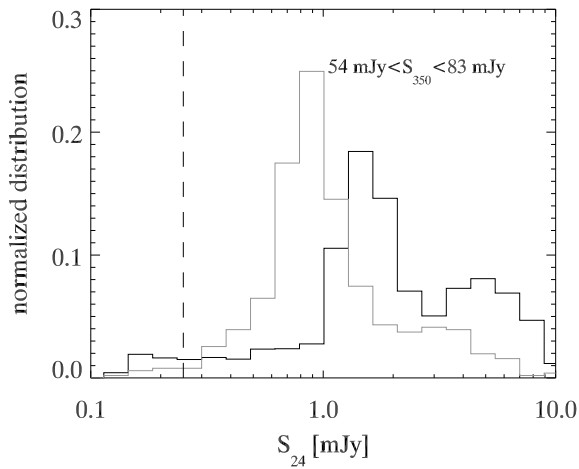}
\caption{\label{fig:miss_prior} Flux density distribution at 24~$\mu$m
  of the 54~mJy~$<S_{350}<$~83~mJy sources. The \citet{Lagache2004} model is
  plotted in \textit{black} and the \citet{Le_Borgne2009} model is
  plotted in \textit{grey}. The \textit{dashed line} represents the
  cut of our catalog at 24~$\mu$m.}
\end{figure}

The $S_{250}/S_{24}$ ($S_{350}/S_{24}$ or $S_{500}/S_{24}$)
  color is not constant, and some sources with a high color ratio
  could have been missed in the prior catalog (especially
  high-redshift starbursts). We used the \citet{Lagache2004} and \citet{Le_Borgne2009} models to estimate the
  fraction of sources missed. We selected the sources in the sub-mm flux
  density bin and computed the 24~$\mu$m flux density distribution (see
  Fig. \ref{fig:miss_prior}). According to the \citet{Lagache2004} model, 99.6\%, 96.4\% and 96.9\% of the sub-mm selected sources$^4$ are brighter than $S_{24}=$~250~$\mu$Jy for a selection at 250~$\mu$m, 350~$\mu$m and 500~$\mu$m, respectively The \citet{Le_Borgne2009} model gives
  99.8\%, 98.3\% and 95.0\%, respectively.
  
\subsection{Photometric accuracy}

The photometric accuracy was estimated with Monte-Carlo artificial
sources. We added five sources of a given flux at random positions on the
original map and add them to the 24~$\mu$m catalog. We then performed a
PSF fitting and compared the input and output flux. We did this 100 times per
tested flux for 10 flux densities (between 10 and 100~mJy). In this simulation we assumed that the position of the sources is exactly known. It is a reasonable hypothesis due to the 24~$\mu$m PSF FWHM (6.6'') compared to the BLAST one (36'' at 250~$\mu$m).

We did not detect any boosting effect for faint flux densities as
expected in this case of detection using a prior. For a blind extraction there is a bias of selection toward sources located on peaks of (instrumental or confusion) noise. This is not the case for an extraction using a prior, for which the selection is performed at another wavelength.

The scatter of output flux densities is the same for all the input
flux densities. We found a photometric noise $\sigma_S$ of 21.5~mJy, 18.3~mJy and 16.6~mJy at 250~$\mu$m, 350~$\mu$m and 500~$\mu$m, respectively. It is slightly lower than for the blind extraction, for which the position of source is not initially known.

\subsection{Estimation of the number counts}

 From the catalog described in Sect. \ref{section:prior_cat} we give an
  estimation of the submillimeter number counts at flux densities
  fainter than reached by the blind-extracted catalog. We cut the
  prior catalog at 3~$\sigma_S$, corresponding to 64~mJy, 54~mJy and 49~mJy
  at 250~$\mu$m, 350~$\mu$m and 500~$\mu$m, respectively. We worked in a single flux density
  bin, which is defined to be between this value and the cut of the blind-extracted
  catalog\footnote{the bins are defined as 64 to 97 at 250~$\mu$m, 54 to 83 at 350~$\mu$m and 49 to 76 at 500~$\mu$m}. There is no flux boosting effect, but we needed to correct
  the Eddington bias. The completeness could not be defined in the same way as for the
  blind extraction, because the selection was performed at another
  wavelength. We thus cannot suppose power-law counts, because the selection function is unknown and the distribution of the extracted sources cannot be computed.
  
The Eddington bias was estimated with another method. We took the sub-mm flux of each of the sources selected at 24~$\mu$m and computed how many sources lie in our count bin. We added a Gaussian noise $\sigma_S$ to the flux of each source to simulate the photometric errors. We computed the number of sources in the counts bin for the new fluxes. We then compute the mean of the ratio between the input and output number of sources in the selected bin for 1000 realizations. The estimated ratios are 0.42, 0.35 and 0.21 at 250~$\mu$m, 350~$\mu$m and 500~$\mu$m, respectively. These low values indicate that on average the photometric noise introduces an excess of faint sources in our flux bin. This effect is strong because of the steep slope of the number counts, implying more fainter sources than brighter sources. The results are interpreted in the Sect. \ref{section:counts_int}.
 
 \subsection{Sub-mm/24 color}
 
 \label{section:color_prior}
 
In this part we work only on S $>$ 5 $\sigma_S$ sources of the catalog described in Sect. \ref{section:prior_cat}  to avoid bias due to the Eddington bias in our selection. At 250~$\mu$m, we have two sources verifying this criterion with a S$_{250}$/S$_{24}$ color of 16 and 60. No sources are brighter than 5 $\sigma_S$ at larger wavelengths. For this cut in flux (S$_{250}>$~5~$\sigma_S$), the \citet{Lagache2004} and \citet{Le_Borgne2009} models predict a mean S$_{250}$/S$_{24}$ color of 39 and 41, respectively. The two models predict a mean redshift of 0.8 for this selection, and the K-correction effect explains these high colors.
 
 \begin{table*}
\centering
\begin{tabular}{|l|lll|r|rr|l|}
\hline
Wavelength & S$_{mean}$ & S$_{min}$ & S$_{max}$ & N$_{sources}$ & S$^{2.5}$.dN/dS & $\sigma_{S^{2.5}.dN/dS}$ & Method \\
\hline
$\mu$m & \multicolumn{3}{|c|}{mJy} & galaxies & \multicolumn{2}{|c|}{$gal.Jy^{1.5}.sr^{-1}$} & \\ 
\hline
250 &      79 &       64 &       97 &       26 &     4451 &     1203 &       Prior \\
250 &        116 &       97 &      140 &       5 &     4529 &     2090 &       Deep \\
250 &        168 &      140 &      203 &       3 &     4040 &     2377 &       Deep \\
250 &        261 &      203 &      336 &       34 &     2987 &      784 &       Shallow \\
250 &        430 &      336 &      552 &       5 &     1023 &      479 &       Shallow \\
250 &        708 &      552 &      910 &       1 &      445 &      449 &       Shallow \\
250 &       1168 &      910 &     1500 &       2 &     1939 &     1401 &       Shallow \\
\hline	
350 &      67 &       54 &       83 &       17 &     1913 &      630 &       Prior \\
350 &      115 &       83 &      161 &       2 &      955 &      687 &       Deep \\
350 &     223 &      161 &      310 &       16 &      854 &      299 &       Shallow \\
 350 &    431 &      310 &      600 &       2 &      314 &      228 &       Shallow \\
 \hline
500 &       61 &       49 &       76 &       7 &      388 &      178 &       Prior \\
 500 &     99 &       76 &      131 &       1 &      443 &      448 &       Deep \\
  500 &   185 &      131 &      262 &       4 &      129 &       99 &       Shallow \\
 \hline
\end{tabular}
\caption{\label{tab:nc} Number counts deduced from source extraction. The not normalized counts can be obtained dividing the S$^{2.5}$.dN/dS column by the S$_{mean}$ column.}
\end{table*}


\section{Non-resolved source counts by stacking analysis}

\subsection{Method}

In order to probe the non-resolved source counts, we used same method
as \citet{Bethermin2010a}, i.e. the stacking analysis applied to number
counts (hereafter ``stacking counts''). We first measured the mean flux at
250~$\mu$m, 350~$\mu$m or 500~$\mu$m as a function of the 24 $\mu$m
flux ($\overline{S_{250,~350~or~500}} = f(S_{24})$). This measurement
was performed by stacking in several $S_{24}$ bins. We used the \citet{Bethermin2010a} catalog at 24~$\mu$m of the FIDEL survey. It is deeper than the SWIRE one used in Sect. \ref{section:prior}, but covers a smaller area (0.25 deg$^2$). The photometry of stacked images was performed with the PSF fitting method
(Sect. \ref{section:psffit}), and the uncertainties on the mean flux are
computed with a bootstrap method \citep{Bavouzet_thesis}. We then
computed the counts in the sub-mm domain with the following formula:
\begin{equation}
\frac{dN}{dS_{submm}}\biggl | _{S_{submm} = f(S_{24})} = \frac{dN}{dS_{24}}\biggl | _{S_{24}} \big /   \frac{dS_{submm}}{dS_{24}}\biggl | _{S_{24}}.
\label{eq:stacking_counts}
\end{equation}

We show in appendix \ref{section:clusstacking} that the
 clustering effect can be neglected. The results are given in Table \ref{tab:sc} and are plotted in Fig. \ref{fig:nc}.
 
 \subsection{Validity of the stacking analysis in the sub-mm range}
 
There are 1.8, 2.4 and 4.5 S$_{24}>$70~$\mu$Jy sources per BLAST beam at 250~$\mu$m, 350~$\mu$m and 500~$\mu$m, respectively. We thus stacked several sources per beam. \citet{Bethermin2010a} showed that the stacking analysis is valid at 160~$\mu$m in the \textit{Spitzer} data, where the size of the beam is similar to the BLAST one.

\begin{figure}
\centering
\includegraphics{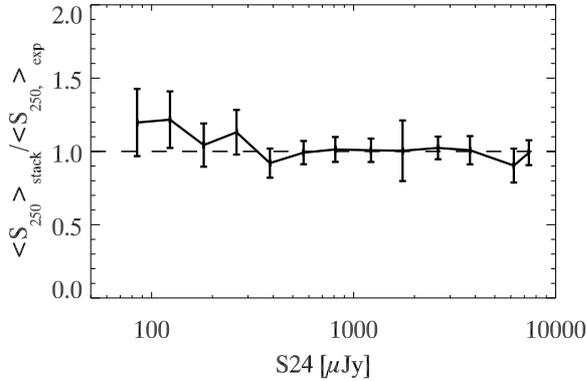}
\caption{\label{fig:valid_stacking} Ratio between the mean flux density at 250~$\mu$m found by the stacking analysis and the expected flux for different S$_{24}$ bins. It is based on a \citet{Fernandez-Conde2008} simulation of a 0.25~deg$^2$ field with a noise and PSF similar to the BLAST deep region.}
\end{figure}

To test the validity of the stacking analysis in the BLAST data from a \textit{Spitzer} 24~$\mu$m catalog, we generated a simulation of a 0.25~deg$^2$ with a Gaussian noise at the same level as for the real map and with source clustering, following \citet{Fernandez-Conde2008}. We stacked the 24~$\mu$m simulated sources per flux bin in the BLAST simulated maps. We measured the mean BLAST flux for each 24~$\mu$m bin with the same method as applied on the real data. At the same time we computed the mean sub-mm flux for the same selection from the mock catalog associated to the simulation. We finally compared the mean BLAST fluxes measured by stacking with those directly derived from the mock catalog to estimate the possible biases (see Fig. \ref{fig:valid_stacking}). The stacking measurements and expected values agree within the error bars. We notice a weak trend of overestimation of the stacked fluxes at low flux density (S$_{24}<$~200~$\mu$Jy) however, but it is still within the error bars. We can thus stack 24~$\mu$m \textit{Spitzer} sources in the BLAST map.

 \subsection{Mean 24~$\mu$m to sub-mm color deduced by stacking analysis}
 
\begin{figure}
\centering
\includegraphics{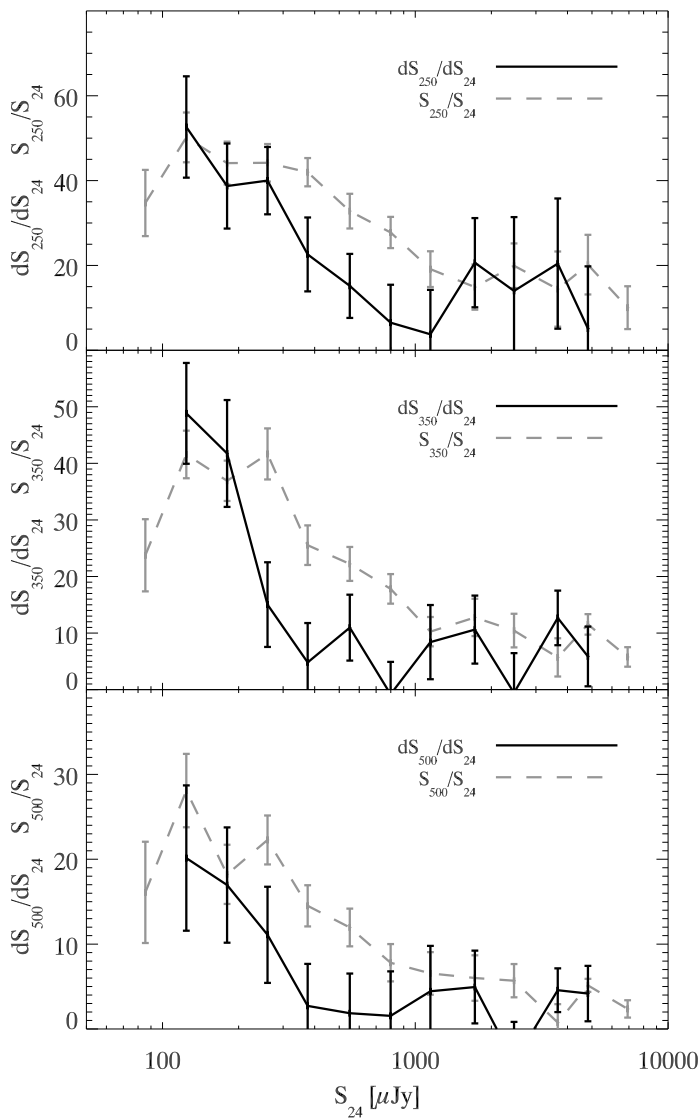}
\caption{\label{fig:color} \textit{Black solid line:} $dS_{submm}/dS_{24}$ as a function of $S_{24}$. \textit{Grey dashed line:}  $S_{submm}/S_{24}$ color as a function of $S_{24}$.}
\end{figure}

The stacking analysis allowed to measure the mean 24~$\mu$m to sub-mm colors of undetected sub-mm galaxies. These colors depends on the SED of galaxies (or K-correction) and the redshift distribution in a degenerate way. The $S_{submm}/S_{24}$ color and $dS_{submm}/dS_{24}$ as a function of $S_{24}$ are plotted in Fig. \ref{fig:color}. 

The colors are higher for the fainter 24~$\mu$m flux ($S_{24} <$ 100~$\mu$Jy). This behavior agrees with the model expectations: the faint sources at 24~$\mu$m lie at a higher mean redshift than the brighter ones. Due to the K-correction, the high-redshift sources have a brighter sub-mm/24 color than local ones.

The colors found by the stacking analysis are lower than those obtained by an extraction at 250~$\mu$m (Sect. \ref{section:color_prior}). It is an effect of selection. The mid-infrared is less affected by the K-correction than the sub-mm , and a selection at this wavelength selects lower redshift objects. We thus see lower colors because of the position of the SED peak (around 100~$\mu$m rest-frame).

We also investigated the evolution of the derivative $dS_{submm}/dS_{24}$ as a function of $S_{24}$, which explicits how the observed sub-mm flux increases with the 24~$\mu$m flux densities. At high 24~$\mu$m flux densities ($S_{24}>$~400~$\mu$Jy) the derivative is almost constant and small ($<$20 and compatible with zero), meaning that the observed sub-mm flux density does not vary much with $S_{24}$. For these flux bins we select only local sources and do not expect a strong evolution of the color. At fainter 24~$\mu$m flux densities the observed decrease can be explained by redshift and K-correction effects, as above. 

The color in the faintest 24~$\mu$m flux density bin (70 to 102 $\mu$Jy) is slightly fainter than in the neighboring points. It can be due to the slight incompleteness of the 24~$\mu$m catalog (about 15\%), which varies spatially across the field: the sources close to the brightest sources at 24~$\mu$m are hardly extracted. The consequence is a bias to the lower surface density regions, leading to a slight underestimation of the stacked flux measurement.

\subsection{Accuracy of the stacking counts method on BLAST with a \textit{Spitzer} 24 $\mu$m prior}

\citet{Bethermin2010a} showed that the stacking counts could be biased:
the color of sources can vary a lot as a function of the redshift. The assumption of a single color for a given S$_{24}$ is not totally realistic and explains some biases. We used two simulated catalogs (containing for each source $S_{24}$,
$S_{250}$, $S_{350}$ and $S_{500}$) to estimate this effect: a first one
based on the \citet{Lagache2004} model that covered 20$\times$2.9 deg$^2$ and a second one based on the \citet{Le_Borgne2009} model and that covered 10~deg$^2$. The large size of these simulations allows us to neglect cosmic
variance.

In order to compute the stacking counts, we first computed the counts
at 24 $\mu$m from the mock catalog. Then we computed the mean
$S_{250,~350~or~500}$ flux density (directly in the catalog) in several $S_{24}$ bins to
simulate a stacking. We finally applied the Eq. \ref{eq:stacking_counts} to compute stacking counts at the BLAST
wavelengths.

\begin{figure*}
\centering
\includegraphics{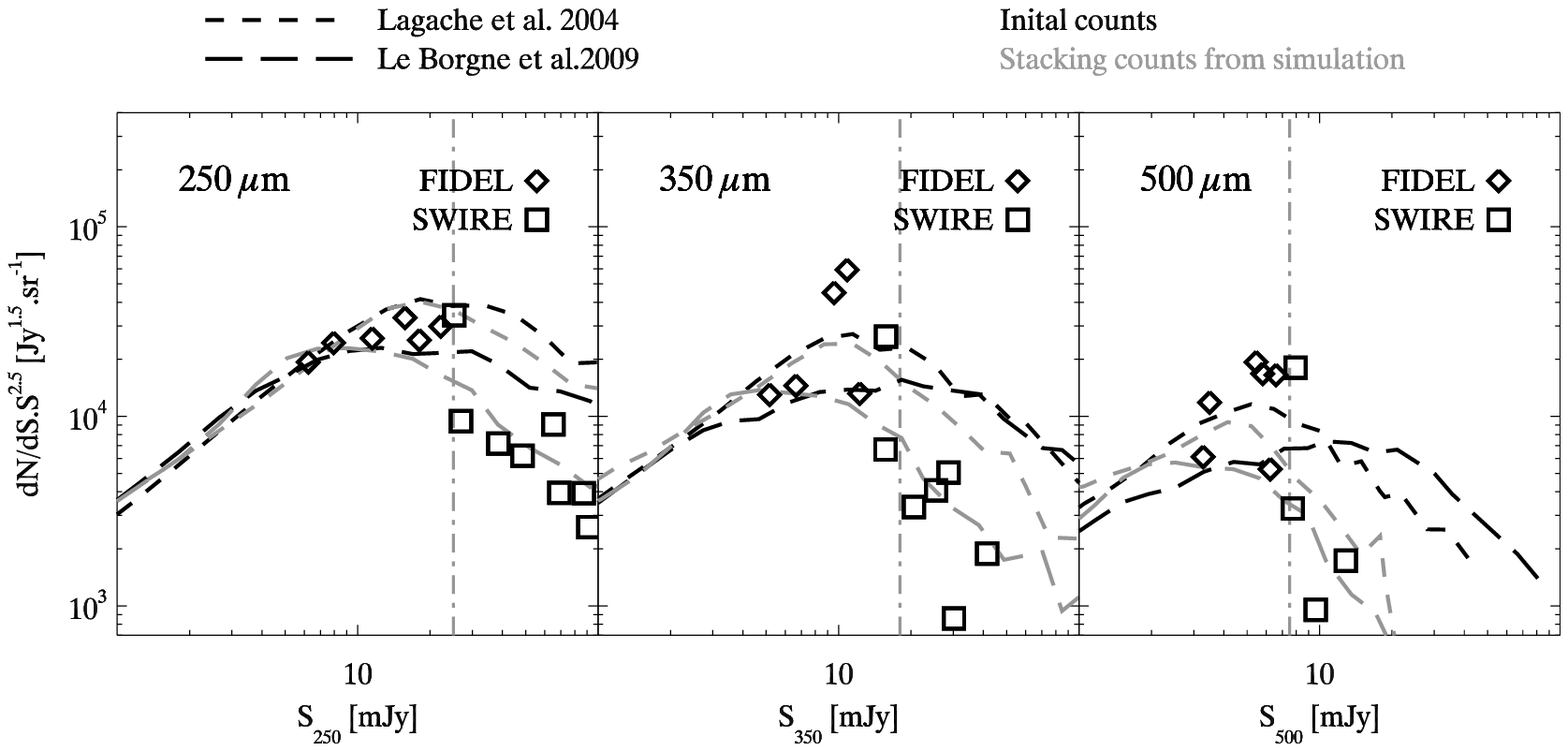}
\caption{\label{fig:comp_stacking_counts} Number counts at BLAST
  wavelengths coming from the data (\textit{points}) and the models (\textit{lines}). \textit{Short dashed line}: initial \textit{(black)} and stacking \textit{(grey)}
  counts from the \citet{Lagache2004} mock catalog; \textit{long dashed
    line}: initial \textit{(black)} and stacking \textit{(grey)}
  counts from \citet{Le_Borgne2009}; \textit{diamond}: stacking
  counts built with the FIDEL catalog; \textit{square}: stacking
  counts built with the SWIRE catalog; \textit{grey vertical dot-dash line}:
  flux cut for stacking counts.}
\includegraphics{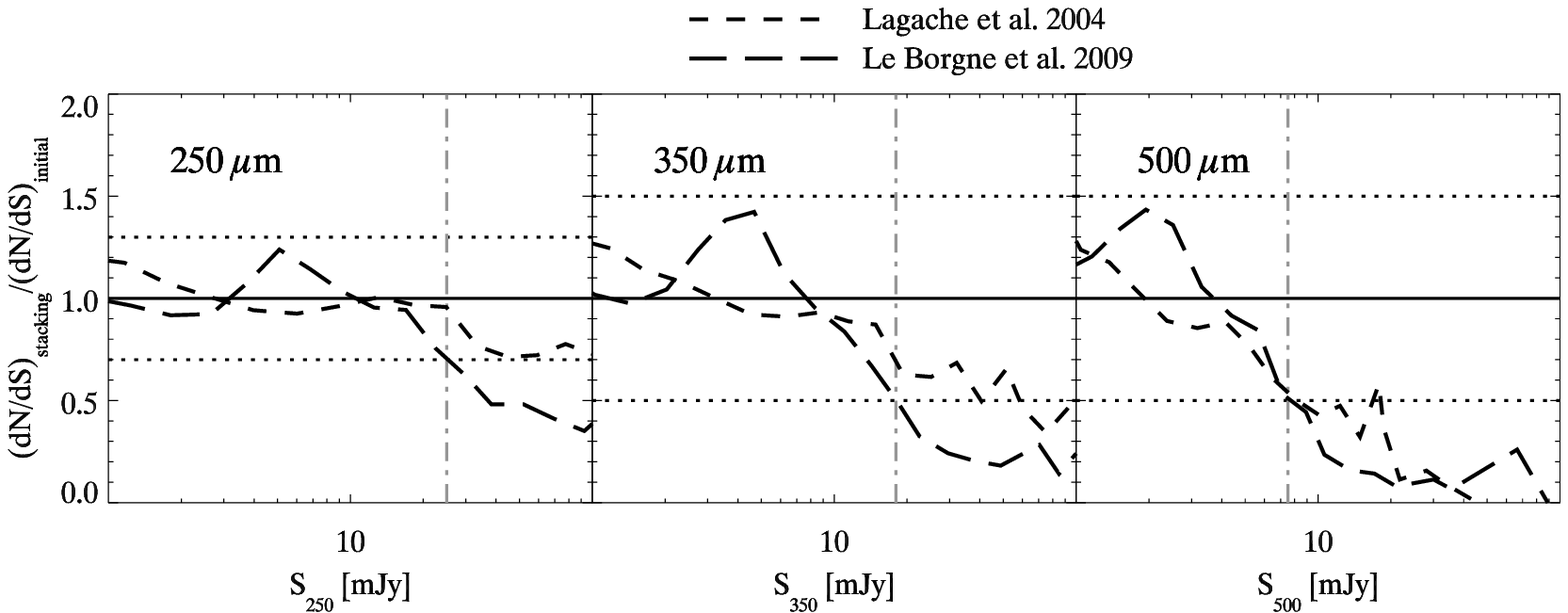}
\caption{\label{fig:valid_stacking_counts} Ratio between stacking
  counts and initial counts for two mock catalogs; \textit{short dashed line}: \citet{Lagache2004} catalog; \textit{long dashed line}:
  \citet{Le_Borgne2009}; \textit{grey vertical dot-dash line}: flux cut for
  stacking counts; \textit{horizontal dot line}: estimation of the
  uncertainty intrinsic to the stacking method.}
\end{figure*}

The ratio between the stacking counts and the initial counts is plotted
in Fig. \ref{fig:valid_stacking_counts} for the two mock
catalogs. Between 1~mJy and 10~mJy we observe an oscillating bias. This
bias is less than 30\% at 250~$\mu$m and 50\% at other
wavelengths. When the flux becomes brighter than 25~mJy at 250~$\mu$m
(18~mJy at 350~$\mu$m and 7.5~mJy at 350~$\mu$m), we begin to strongly
underestimate the counts. The analysis of real data also shows a very
strong decrease in the counts around the same fluxes (see Fig.
\ref{fig:comp_stacking_counts}). Consequently, we cut our stacking
analysis at these fluxes and we applied an additional uncertainty to
the stacking counts of 30\% at 250~$\mu$m (50\% at 350~$\mu$m and
500~$\mu$m).

\begin{figure}
\centering
\includegraphics{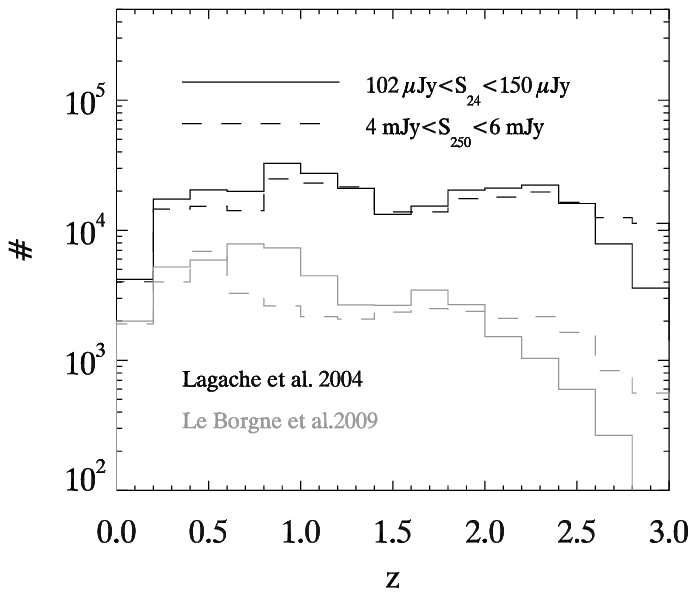}
\caption{\label{fig:nz_faint} \textit{Solid line:} Distribution in
  redshift of the sources with 102~$\mu$Jy~$<S_{24}<$~150~$\mu$Jy for the mock
  catalogs generated with the \citet{Lagache2004} (\textit{black}) and
  the \citet{Le_Borgne2009} (\textit{grey}) models; \textit{Dashed line:}
  Distribution in redshift of the sources with 4~mJy~$<S_{250}<$~6~mJy
  (determined using the mean 250/24 color).}
\includegraphics{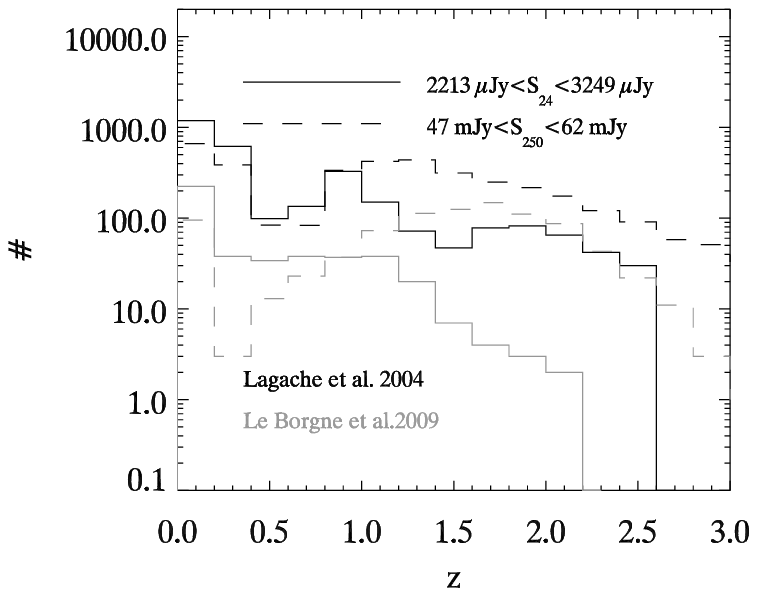}
\caption{\label{fig:nz_bright} \textit{Solid line:} Distribution in
  redshift of the sources with 2213~$\mu$Jy~$<S_{24}<$~3249~$\mu$Jy for
  the mock catalogs generated with the \citet{Lagache2004} (\textit{black})
  and \citet{Le_Borgne2009} (\textit{grey}) models; \textit{Dashed
    line:} Distribution in redshift of sources with
  47~mJy~$<S_{250}<$~62~mJy (determined using the mean 250/24 color).}
\end{figure}

Using the 24 $\mu$m observations as a prior to stack in the BLAST bands seems to give less accurate results than in the \textit{Spitzer} MIPS bands. For a given $S_{24}$ flux, the sub-mm emission can vary a
lot as a function of the redshift. But the simulations shows that
this method works for faint flux densities. It is due to the redshift
selection which is similar for faint flux densities (see Fig.
\ref{fig:nz_faint}) and very different at higher flux densities (see
Fig. \ref{fig:nz_bright}). For example, $S_{24} \sim $100$~\mu$Jy
sources are distributed around $z=1.5$ with a broad dispersion in
redshift. S$_{350} \sim$~4~mJy (based on averaged colors, 4~mJy at
350~$\mu$m corresponds to S$_{24} \sim$~100~$\mu$Jy) sources have quite a similar redshift distribution except an excess for $z>2.6$. At higher flux densities (around 2~mJy at 24~$\mu$m) the
distribution is very different. The majority of the 24~$\mu$m-selected
sources lies at $z<1$ and the distribution of 350~$\mu$m-selected
sources peaks at $z\sim$1.5. Another possible explanation is that fainter sources lies near z=1 and are thus selected at the 12~$\mu$m rest-frame, which is a very good estimator of the infrared bolometric luminosity according to \citet{Spinoglio1995}. 

In order to limit the scatter of the sub-mm/24 color, we tried to cut our
sample into two redshift boxes following the \citet{Devlin2009} IRAC
color criterion ([3.6]-[4.5]=0.068([5.8]-[8.0])-0.075). But we had not enough signal in the stacked images to
perform the analysis.

\begin{table}
\centering
\begin{tabular}{|l|ll|rr|}
\hline
Wavelength & S & $\sigma_S$ & S$^{2.5}$.dN/dS & $\sigma_{S^{2.5}.dN/dS}$ \\
\hline
$\mu$m & \multicolumn{2}{|c|}{mJy} & \multicolumn{2}{|c|}{$gal.Jy^{1.5}.sr^{-1}$} \\ 
\hline
 250 &        6.2 &          0.7 &       19313 &        7892 \\
 250 &           7.9 &          0.9 &       24440 &       10466 \\
 250 &          11.5 &          1.2 &       25816 &       10236 \\
   250 &        15.7 &          1.3 &       33131 &       17213 \\
      250 &     18.1 &          2.3 &       25232 &       15428 \\
      250 &     22.2 &          2.9 &       29831 &       42448 \\
       \hline     
       350 &  5.2 &          0.5 &       13007. &        7343. \\
350 &          6.6 &          0.6 &       14519. &        8434. \\
350 &         10.8 &          1.2 &       59314. &       43418. \\
350 &          9.6 &          1.3 &       44944. &       69505. \\
350 &         12.2 &          1.6 &       13200. &       10044. \\    
       \hline
500 &         3.5 &          0.5 &       11842. &        8134. \\
 500 &        3.3 &          0.6 &        6115. &        4112. \\
500 &         5.8 &          0.8 &       16789. &       12498. \\
500 &         5.4 &          0.9 &       19338. &       36659. \\
500 &         6.6 &          1.2 &       16526. &       42476. \\
 500 &        6.2 &          1.8 &        5263. &       18087. \\    
\hline	
\end{tabular}
\caption{\label{tab:sc} Number counts deduced from stacking. The not normalized counts can be obtained dividing the S$^{2.5}$.dN/dS column by the S column.}
\end{table}

\section{Interpretation}
\label{section:counts_int}

\subsection{Contribution to the CIB}

We integrated our counts assuming power-law behavior between our points. Our points are not independent (especially the stacking counts), and we thus combined errors linearly. The contribution of the individually detected sources (S$_{250}$~$>$~64~mJy, S$_{350}$~$>$~54~mJy, S$_{500}$~$>$~49~mJy) is then 0.24$^{+0.18}_{-0.13}$~nW.m$^2$.sr$^{-1}$, 0.06$^{+0.05}_{-0.04}$~nW.m$^2$.sr$^{-1}$ and 0.01$^{+0.01}_{-0.01}$~nW.m$^2$.sr$^{-1}$ at 250~$\mu$m, 350~$\mu$m and 500~$\mu$m, respectively. Considering the total CIB level of \citet{Fixsen1998} (FIRAS absolute measurement), we resolved directly only 2.3\%, 1.1\% and 0.4\% at 250~$\mu$m, 350~$\mu$m and 500~$\mu$m, respectively.

The populations probed by the stacking counts (S$_{250}$~$>$~6.2~mJy, S$_{350}$~$>$~5.2~mJy, S$_{500}$~$>$~3.5~mJy) emit 5.0$^{+2.5}_{-2.6}$~nW.m$^2$.sr$^{-1}$, 2.8$^{+1.8}_{-2.0}$~nW.m$^2$.sr$^{-1}$ and 1.4$^{+2.1}_{-1.3}$~nW.m$^2$.sr$^{-1}$ at 250~$\mu$m, 350~$\mu$m and 500~$\mu$m, respectively. This corresponds to about 50\% of the CIB at these three wavelengths.

\subsection{Comparison with \citet{Patanchon2009}}

The agreement between our resolved counts built from the catalogs and
the P(D) analysis of \citet{Patanchon2009} is excellent
(Fig. \ref{fig:nc}). We
confirm the efficiency of the P(D) analysis to recover number counts without extracting sources. The stacking counts probe the flux densities between 6~mJy and 25~mJy at 250~$\mu$m (between 5~mJy and
  13~mJy at 350~$\mu$m and 3~mJy and 7~mJy at 500~$\mu$m). In this range there is only one P(D) point. At the three BLAST wavelengths the P(D) points agree with our stacking counts
(Fig. \ref{fig:nc}). Our results thus confirm the measurement of \citet{Patanchon2009} and give
a better sampling in flux.

\subsection{Comparison with ground-based observations}

We compared our results with sub-mm ground-based observations of
SHARC. \citet{Khan2007} estimated a density of
S$_{350}$~$>$~13~mJy sources of 0.84$_{-0.61}^{+1.39}$
arcmin$^{-2}$. For the same cut, we found 0.26$\pm$0.13 arcmin$^{-1}$, which agrees with their work.
Our measurement (175$\pm$75~sources.deg$^{-2}$ brighter than 25~mJy) agrees also with that of \citet{Coppin2008} ones
at the same wavelength (200-500~sources.deg$^{-2}$ brighter than 25~mJy).

We also compared our results at 500~$\mu$m with the SCUBA ones at
450~$\mu$m. \citet{Borys2003} find 140$_{-90}^{+140}$ gal.deg$^{-2}$
for S$_{450}>$100~mJy. We found 1.2$\pm$1.0
gal.deg$^{-2}$. We significantly disagree with them. \citet{Borys2003} claim 5 4-$\sigma$ detections in a 0.046 deg$^2$ field in the \textit{Hubble} deep field north (HDFN). These five sources are brighter than 100~mJy. We find no source brighter than 100~mJy in a 0.45~deg$^2$ field at 350~$\mu$m nor at 500~$\mu$m. The cosmic variance alone thus cannot explain this difference. A possible explanation is that they underestimated the noise level and their detections are dominated by spurious sources. It could also be due to a calibration shift (by more than a factor 2). The observation of the HDFN by \textit{Herschel} will allow us to determine whether that these bright sources might be spurious detections.

We also compared our results with the estimations based on lensed sources at 450 $\mu$m
with SCUBA \citep{Smail2002,Knudsen2006}. For example,
\citet{Knudsen2006} find 2000-50000 sources.deg$^2$ brighter than 6~mJy. It agrees with our 3500$^{+7700}_{-3400}$~sources.deg$^2$.

\subsection{Comparison with the \citet{Lagache2004} and \citet{Le_Borgne2009} models}

At 250~$\mu$m and 350~$\mu$m the measured resolved source counts are
significantly lower (by about a factor of 2) than the
\citet{Lagache2004} and \citet{Le_Borgne2009} models. Nevertheless,
our counts are within the confidence area of
\citet{Le_Borgne2009}. The same effect (models overestimating the
counts) was observed at 160 $\mu$m
\citep{Frayer2009,Bethermin2010a}. It indicates that the galaxies' SED or the luminosity functions
used in both models might have to be revisited. At
500~$\mu$m our counts and both models agree very well, but our uncertainties are large, which renders any discrimination difficult.

Concerning the stacking counts, they agree very well with
the two models. Nevertheless, our uncertainties are larger than 30\%. We
thus cannot check if the disagreement observed between the
\citet{Lagache2004} model and the stacking counts at 160 $\mu$m
\citep{Bethermin2010a} of 30\% at $S_{160}$~=~20~mJy still holds at 250
$\mu$m.

\subsection{Implications for the probed populations and the models}

\begin{figure}
\centering
\includegraphics{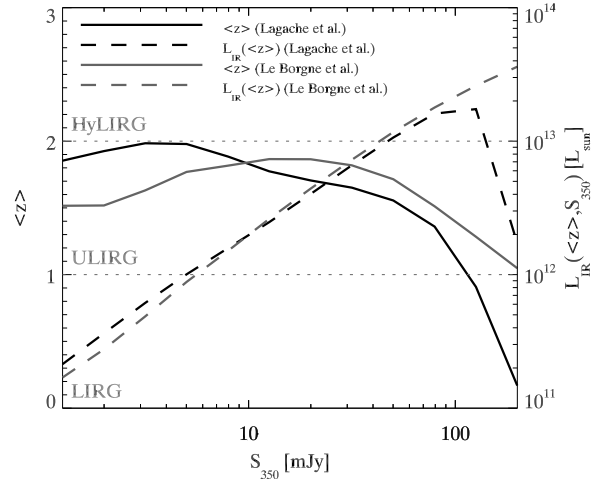}
\caption{\label{fig:z_mean} Mean redshift (\textit{solid line}) of
  sources for different fluxes at 350 $\mu$m for the \citet{Lagache2004}
  (\textit{black}) and \citet{Le_Borgne2009} (\textit{grey}) models and corresponding infrared luminosity defined in Sect. \ref{section:popred} (\textit{dashed line}).}
\end{figure}

\label{section:popred}

We showed that the two models nicely reproduce the sub-mm counts,
especially below 100~mJy. We can thus use them to estimate which
populations are constrained by our counts. For each flux density bin
we computed the mean redshift of the selected galaxies in both models. We then used the SEDs given by the models at that mean redshift  and at that flux bin and derived the infrared bolometric luminosity. The luminosities are
shown in dashed lines in Fig. \ref{fig:z_mean} for 350~$\mu$m as an
example, and the redshift is given in solid lines.

The stacking counts reach 6.2~mJy, 5.3~mJy and 3.5~mJy at 250~$\mu$m, 350~$\mu$m and
500~$\mu$m, respectively. This corresponds to faint ULIRGs ($L_{IR} \approx 1.5 \times
10^{12} L_{\sun}$) around $z=$~1.5, 1.8 and 2.1 at 250~$\mu$m, 350~$\mu$m and
500~$\mu$m, respectively.  Our measurements show that the predicted cold-dust
emissions (between 100~$\mu$m and 200~$\mu$m rest frame) of this
population in the models are believable.

At 250~$\mu$m and 350~$\mu$m the resolved sources
($S_{350}>85$~mJy) are essentially $z \sim 1$ ULIRGs ($L_{IR} > 10^{12} L_{\sun}$) and HyLIRGs ($L_{IR} > 10^{13} L_{\sun}$) according to the models. In \citet{Lagache2004} the local cold-dust sources contribute at very
bright flux ($>$~200mJy). This population is not present in the
\citet{Le_Borgne2009} model. It explains the difference between the
two models for fluxes brighter than 100~mJy at 350 $\mu$m (see Fig.
\ref{fig:z_mean}). At 500~$\mu$m, \citet{Lagache2004} predict that
bright counts are dominated by local cold-dust populations and
\citet{Le_Borgne2009} that they are dominated by medium redshift
HyLIRGs. Nevertheless, there is a disagreement with the observations
for this flux density range, suggesting that there could be less
HyLIRGs than predicted. But these models do not currently include any AGN contribution, which is small except at luminosities higher than 10$^{12}$ L$_\odot$ \citep{Lacy2004,Daddi2007,Valiante2009}.

\section{Conclusion}

Our analysis provides new stacking counts, which can be compared with the \citet{Patanchon2009} P(D) analysis. We have a good agreement between the different methods. Nevertheless, some methods are more efficient in a given flux range.

The blind extraction and the extraction using a prior give a better sampling in flux and slightly smaller error bars. The P(D) analysis uses only the pixel histogram and thus looses the information on the shape of the sources. The blind extraction is a very efficient method for extracting the sources, but lots of corrections must be applied carefully. When the confusion noise totally dominates the instrumental noise, the former must be determined accurately, and the catalog flux limit must take this noise \citep{Dole2003} into account.

Estimating the counts from a catalog built using a prior is a good way to deal with the flux boosting effect. This method is based on assumptions however. We assume that all sources brighter than the flux cut at the studied wavelength are present in the catalog extracted using a prior. We also assume a flux distribution at the studied wavelength for a selection at the prior wavelength to correct for the Eddington bias. Consequently an extraction using a prior must be used in a flux range where the blind extraction is too affected by the flux boosting to be accurately corrected.

P(D) analysis and stacking counts estimate the counts at flux densities below the detection limit. These methods have different advantages. The P(D) analysis fits all the fluxes at the same time, where the stacking analysis flux depth depends on the prior catalog's depth (24 $\mu$m \textit{Spitzer} for example). But the P(D) analysis with a broken power-law model is dependent on the number and the positions of the flux nodes. The uncertainty due to the parameterization was not evaluated by \citet{Patanchon2009}. The stacking counts on the other hand are affected by biases due to the color dispersion of the sources. The more the prior and stacked wavelength are correlated, the less biased are the counts. A way to overcome this bias would be to use a selection of sources (in redshift slices for example), which would reduce the color dispersion, and the induced bias; we did not use this approach here because of a low signal-to-noise ratio.

The stacking and P(D) analysis are both affected by the clustering in different ways. For the stacking analysis this effect depends on the size of the PSF. This effect is small for BLAST and will be smaller for SPIRE. The clustering broadens the pixel histogram. \citet{Patanchon2009} show that it is negligible for BLAST. Clustering will probably be an issue for SPIRE. The cirrus can also affect the P(D) analysis and broaden the peak. \citet{Patanchon2009} use a high-pass filtering that reduces the influence of these large scale structures.

The methods used in this paper will probably be useful to perform the
analysis of the \textit{Herschel} SPIRE data. The very high sensitivity and the large area covered will reduce the uncertainties and  increase the depth of the resolved source counts. Nevertheless, according to the models (e.g. \citet{Le_Borgne2009}), the data will also be quickly confusion-limited and it will be very hard to directly probe the break of the counts. The P(D) analysis of the deepest SPIRE fields will allow us to constrain a model with more flux nodes and to better sample the peak of the normalized differential number counts. The instrumental and confusion noise will be lower, and a stacking analysis per redshift slice will probably be possible. These analyses will give stringent constraints on the model of galaxies and finally on the evolution of the infrared galaxies.

\begin{acknowledgements}
  We warmly acknowledge Guillaume Patanchon for his precious comments
  and discussions. We thank Damien Le Borgne and Guilaine Lagache for
  distributing their model and their comments. We also acknowledge
  Alexandre Beelen and Emeric Le Floc'h for their useful comments. We thank Maxime Follin,
  for his help during his Licence 3 training at the Universit\'e Paris
  Sud 11.  We thank the BLAST team for the well-documented public
  release of their data.
  
  We warmly thank the referee (Steve Willner), who helps a lot to improve the quality of this paper.
  
  This work is based in part on archival data obtained
  with the Spitzer Space Telescope, which is operated by the Jet
  Propulsion Laboratory, California Institute of Technology under a
  contract with NASA. Support for this work was provided by an award
  issued by JPL/Caltech.
\end{acknowledgements}

\bibliographystyle{aa}

\bibliography{biblio}

\begin{appendix}
\section{Effect of clustering on the uncertainties of number counts}

\label{section:cluscounts}

\citet{Bethermin2010a} showed how the clustering is linked with the
uncertainties of the counts. We used the formalism of \citet{Bethermin2010a} to
estimate the effect of the clustering on our BLAST counts. There are
few sources detected at 250~$\mu$m and the BLAST coverage is
inhomogeneous. It is consequently very hard to estimate the clustering
of the resolved population. We thus used the clustering measured at
160~$\mu$m by \cite{Bethermin2010a} and assumed a 250/160
color equal to unity. We then used the same method to compute the uncertainties. We
then compare the uncertainties with and without clustering. Neglecting
the clustering implies an underestimation of the uncertainties on the
counts of 35\% in the 203-336~mJy bin at 250~$\mu$m, and less than 20\% in the
other bins. We can thus suppose a Poissonian behavior, knowing that the Poisson approximation underestimates the error bars for the 203-336~mJy bin at 250~$\mu$m. Nevertheless, our model of clustering at 250~$\mu$m has strong assumptions (single 250/160 color, same clustering at 250~$\mu$m as measured at 160~$\mu$m), and it would be more conservative to update it with \textit{Herschel} clustering measurements.

\section{Effect of clustering on stacking}

\label{section:clusstacking}

\subsection{A formalism to link clustering and stacking}

The clustering can bias the results of a stacking. We present a
formalism based on \citet{Bavouzet_thesis} work.

The expected results for mean stacking of an N non-clustered
populations is
\begin{equation}
M(\theta) = \overline{S_{s}} \times PSF(\theta) + \int_{0}^{\infty} S \frac{dN}{dS} dS,
\label{eq:stacknoclus}
\end{equation}
where M is the map resulting from stacking, $\theta$ the distance to
the center of the cutout image, $\overline{S_{s}}$ the mean flux of
the stacked population. The integral is an approximation because the central source is treated in the first term. This approximation is totally justified in a strongly confused field where the number of sources is enormous. $PSF$ is the instrumental response and
  is supposed to be invariant per rotation ($\theta=0$ corresponds to
  the center of this PSF). $\frac{dN}{dS}$ is the number of the source
per flux unit and per pixel. We assume an absolute calibration. The
integral in the equation \ref{eq:stacknoclus} is equal to the CIB
brightness
\begin{equation}
I_{CIB} = \int_{0}^{\infty} S \frac{dN}{dS} dS.
\label{eq:cib}
\end{equation}
This term is constant for all pixels of the image and corresponds to
a homogeneous background.

The stacked sources can actually be autocorrelated. The probability
density to find a stacked source in a given pixel and another in a
second pixel separated by an angle $\theta$ ($p(\theta)$) is linked
with the angular autocorrelation function ($\omega(\theta)$) by
\begin{equation}
p(\theta) = \rho_{s}^2 (1+\omega(\theta)),
\end{equation}
where $\rho_{s}$ is the number density of the stacked source.

If we assume that there is no correlation with other populations, the
results of the stacking of N autocorrelated sources is
\begin{equation}
M(\theta) = \overline{S_{s}} \times PSF(\theta) + I_{CIB,s} (1+\omega(\theta) ) \ast PSF(\theta) +I_{CIB,ns},
\end{equation}
where $I_{CIB,s}$ and $I_{CIB,ns}$ is the CIB contribution of stacked
and non-stacked sources. If we subtract the constant background of the
image, we find
 \begin{equation}
M(\theta)  = \overline{S_{s}} \times PSF(\theta)  + I_{CIB,s} \times \omega(\theta)  \ast PSF(\theta).
\end{equation}
The second term of this equation corresponds to an excess of
  flux due to clustering. This signal is stronger in the center of the
  stacked image. The central source appears thus brighter than
  expected, because of the contribution due to clustering.

The flux of the central stacked source computed by PSF-fitting photometry is
 \begin{equation}
S_{mes} = \frac{\int \int  M \times PSF d\Omega}{\int \int  PSF^2 d\Omega} = \overline{S_{s}} +  S{_{clus}},
\end{equation}
where $S_{clus}$, the overestimation of flux due to clustering is given by
\begin{equation}
S_{clus} = I_{CIB,s} \times \frac{\int \int \big ( (\omega \ast PSF) \times PSF \big ) d\Omega}{\int \int PSF^2 d\Omega}.
\label{eq:Sclus}
\end{equation}
Basically, the stronger the clustering, the larger the
  bias. In addition, the wider the PSF, the larger the
  overestimation. The stacked signal can be dominated by the
  clustering, if the angular resolution of the instrument is low
  compared to the surface density of the sources (like
  \textit{Planck}, c.f. \citet{Fernandez2010}) or if strongly clustered populations are stacked.

\subsection{Estimation of the bias due to clustering}

The estimation of $S_{clus}$ with Eq. \ref{eq:Sclus} requires particular hypotheses. The stacked population is S$_{24}>$~70~$\mu$Jy sources detected by \textit{Spitzer}. Their contribution to the CIB is 5.8~nW.m$^{-2}$.sr$^{-1}$, 3.4~nW.m$^{-2}$.sr$^{-1}$ and 1.4~nW.m$^{-2}$.sr$^{-1}$ at 250~$\mu$m, 350~$\mu$m and 500~$\mu$m,
respectively (estimated by direct stacking of all the sources). Following the clustering of 24~$\mu$m sources estimated by
\citet{Bethermin2010a}, we suppose the following autocorrelation
function:
\begin{equation}
\omega(\theta) = 2.3\times10^{-4} \times \big(\frac{\theta}{deg}\big)^{-0.8}.
\end{equation}
The excess of flux due to clustering ($S_{clus}$) is then 0.44~mJy, 0.35~mJy
and 0.16~mJy at 250~$\mu$m, 350~$\mu$m and 500~$\mu$m, respectively. This is
significantly lower than the bootstrap uncertainties on these fluxes. We
can thus neglect the clustering.

\subsection{Measurement of the angular correlation function by stacking}

This new formalism provides a simple tool to measure the angular
autocorrelation function (ACF) from a source catalog. This method uses
a map called "density map". One pixel of this map contains the number of
sources centered on it. It is equivalent of a map of unit flux
sources with the $PSF = \delta$ (Dirac distribution). The result of the
stacking is thus
\begin{equation}
M(\theta) = \rho_s \times \delta(\theta) + \rho_s (1+\omega(\theta)).
\end{equation}
The ACF can then be easily computed with
\begin{equation}
\forall \theta \neq 0, \omega(\theta) = \frac{M(\theta)}{\rho_s}-1.
\end{equation}

\end{appendix}

\end{document}